# Compensator-based small animal IMRT enables conformal preclinical dose painting: application to tumor hypoxia

Jordan M. Slagowski[1*], Erik Pearson[2], Rajit Tummala[2], Gage Redler[3], Daniela Olivera Velarde[2], Boris Epel[2], Howard J. Halpern[2], Bulent Aydogan[2]

1. Department of Human Oncology, University of Wisconsin-Madison, Madison WI, USA
2. Department of Radiation and Cellular Oncology, The University of Chicago, Chicago IL, USA
3. Department of Radiation Oncology, Moffitt Cancer Center, Tampa, FL, USA

*Corresponding author. E-mail address: slagowski@wisc.edu

**ABSTRACT:** Techniques for preclinical intensity modulated radiation therapy are being developed to improve translation by replicating the clinical paradigm. This study presents the first treatment planning comparison between small animal IMRT (SA-IMRT) and three-dimensional conformal radiotherapy (CRT) in a model application, oxygen-guided dose painting of tumor hypoxia, using actual mouse data. A novel compensator-based platform was employed to generate SA-IMRT and CRT plans with 2-15 beam angles for seventeen mice with fibrosarcoma tumors. The whole tumor received a dose of 22.5 Gy, with a simultaneous integrated boost of 13 Gy to hypoxic voxels identified via electron paramagnetic resonance imaging. Plan quality was assessed using the Paddick conformity index (CI), uniformity, and dose volume histograms. For 3-angles, SA-IMRT yielded significantly improved dose conformity (median hypoxic CI =0.45 versus 0.17), tumor dose uniformity (11.0% versus 14.3%), and dosimetric spread between boost and non-boost targets (D50% difference = 13.0 Gy [ideal], 13.1 Gy [SA-IMRT], 7. 3 Gy [CRT]). No significant improvement in CI was associated with >3 beam angles (Wilcoxon signed-rank test, $p < 0.05$). This study demonstrates that SA-IMRT provides significant improvements in radiation plan quality and yields dose distributions that more closely mimic the clinical setting relative to current CRT approaches.

## 1. INTRODUCTION

Tumor hypoxia is a source of radiation resistance and an eventual cause of treatment failure that has been known for more than a century.[1–6] Oxygen-guided dose painting, defined as the administration of a locally increased prescription dose (i.e. boost) to a hypoxic tumor sub-volume defined by a tumor oxygen threshold, or a prescription dose scaled to tumor oxygen levels, has been proposed as a method to optimally deliver dose to radioresistant tumor regions and improve treatment outcomes.[7–9] Molecular imaging technologies and small animal tumor models exist and could be applied to probe tumor oxygen concentrations and optimize radiation treatments to hypoxic sub-volumes in preclinical experiments.[9,10] Unfortunately, preclinical radiotherapy experiments have traditionally been limited to using broad beams with large fields and untargeted irradiations to animal bodies or tumors.[11,12] These commonly available systems, generically termed cabinet irradiators, cannot precisely target the potentially small and often complex distributions of hypoxia in murine tumors. In some instances, blocking has been performed to partially irradiate animal organs such as hemithorax or head and neck,[13–16] but, for the most part, technologies to administer precise and conformal radiotherapy in the preclinical context do not exist or are not in common use. This is one potential contributing factor to the recognized poor translation of preclinical results to clinical practice.[17,18] Recently, image guided small animal irradiators have become available that provide improved spatial accuracy and targeting using open fixed field apertures.[19,20] These techniques present an opportunity to shift from using broad untargeted radiation beams to precise preclinical irradiation techniques that more closely mimic clinical technologies. Nonetheless, these advanced and commercially available preclinical irradiators are limited to delivering radiation with open fixed-shape apertures. Although solutions for flexible rectangular field shaping are provided in the form of dynamic collimators by multiple vendors, *concave* dose distributions generally cannot be generated due to a lack of methodology for fluence field modulation.

This is a major limitation as these systems cannot perform dose painting to selectively boost dose to the complex tumor sub-volume targets encountered in hypoxic tumor models. This is also restrictive as data suggests that outcomes could be improved with more conformal irradiation. As one example, *Halpern et al*. investigated the use of electron paramagnetic resonance imaging (EPRI) coupled with an image-guided conformal irradiator to identify and boost hypoxic tumor subregions in mouse fibrosarcoma.[21–24] Injection of an oxygen spin probe into the tail vein of the mouse coupled with EPRI allows tumor oxygen concentrations (pO2) to be imaged with 1-3 mmHg resolution.[21] In an initial set of experiments, hypoxic voxels (defined as pO2 ≤ 10 mmHg) were boosted to a 90% tumor control dose with a standard spherical field aperture after administering an approximately uniform 50% tumor control dose to the whole tumor volume independent of oxygen concentration.[25] Despite approximately 85% of the hypoxic voxels receiving the boost dose, no significant outcome improvement was observed (Figs. S4 and S7 of Ref. 24).[24] In a second set of experiments, conformity of the boost dose to the hypoxic subregions was improved by 3D printing animal/tumor-specific conformal field apertures with a tungsten loaded PLA filament material for two opposed treatment beams.[24] This has been enhanced with data from two more syngeneic tumor models.[26] The use of animal-specific conformal apertures increased the percentage of hypoxic voxels covered by the boost dose. A statistically significant (p=0.04) improvement in clonogenic tumor control was observed in the group of animals receiving a boost dose to the hypoxic subregions relative to the control group, which received an equivalent boost dose to well-oxygenated tumor regions of approximately the same total volume. These are the first significant demonstrations of the benefit from treating hypoxic tumor relative to well oxygenated tumor in a mammalian malignancy.

While these initial results are encouraging, several limitations remain due to technological limitations of the available preclinical irradiators. First, it is difficult to simultaneously administer conformal and homogenous radiation doses in a preclinical setting. This is important as small dose variations (e.g. 10%) may produce large differences in biological responses (e.g. 90%).[27] Second, small disconnected hypoxic tumor subregions may be underdosed using conformal apertures, or conversely, large dose spill may be required to adequately cover all hypoxic voxels. And third, without an ability to modulate the beam intensity and produce concave dose distributions, it is impossible to deliver an equivalent integral dose to well-oxygenated tumor regions compared with the hypoxic boost regions, which may bias results.

To address these and other limitations, several groups have recently proposed systems to perform intensity modulated radiation therapy in small animals (SA-IMRT). These include the use of a static miniature pencil beam coupled with a precision-controlled motion stage for animal translation,[28] a sparse orthogonal collimator system,[29,30] binary multi-leaf collimators,[31] and custom molds filled with high atomic number powders for preferential beam attenuation.[32] Our group has previously described and validated a novel technique to deliver conformal IMRT treatments to small animals using 3D printed animal-specific compensators.[33] A detailed discussion of the advantages and disadvantages of the SA-IMRT systems under development is outside the scope of this work.

While technologies for SA-IMRT continue to quickly advance, no studies have been performed with real animal data to determine whether SA-IMRT might provide statistically significant improvements in preclinical radiation plan quality relative to 3D conformal techniques. Thus, the purpose of this work was to perform the first evaluation of preclinical IMRT versus 3D conformal radiation therapy (CRT) to selectively target (i.e., dose paint) tumor hypoxia in a fibrosarcoma mouse model with an SIB approach. A sub aim of the study was to determine the optimal number of beam angles required for SA-IMRT using real mouse data and spatial distributions of tumor hypoxia identified on EPRI.

## 2. MATERIALS AND METHODS

### 2.A. SA-IMRT platform

The system for SA-IMRT developed by our group is shown in Figure 1 and consists of an Xrad225Cx conformal irradiator with onboard cone beam CT (Precision X-ray, North Branford, CT), an inverse treatment planning system (TPS), 3D printed animal-specific compensators for radiation beam intensity modulation, and a custom brass collimator that secures the beam compensators.[33] The irradiator has a 225 kVp photon treatment beam that was modeled in the open-source MatRad TPS.[34] MatRad uses a pencil beam dose calculation algorithm that was commissioned by curve fitting dose kernels generated using the BEAMnrc/EGSnrc Monte Carlo package via DOS-XYZnrc.[35] The beam model was validated by comparing MatRad computed percentage depth dose curves and beam profiles for circular fields of varying diameter versus measurements performed with EBT3 Gafchromic film in a solid water phantom.[33] Absolute dose calibration was performed following TG-61 guidelines.[36] CT number to electron density conversion was performed by linear interpolation of a seven-point calibration curve produced from measurements acquired with inserts of known materials spanning air to cortical bone.

The TPS accepts a CT image with defined target and organ-at-risk structures and outputs the discrete 2D fluence patterns for a set number of beam angles that minimize a weighted piece-wise quadratic cost function defining target coverage and normal tissue avoidance goals. Each optimized fluence pattern is then converted to an STL file defining a 3D compensator model with thickness of copper doped PLA (CuPLA, 80% Cu, 20% PLA) filament material varying spatially to modulate intensity at each sub-section, or bixel, of a given beam. The CuPLA thickness is determined using an empirical fit of measured beam transmission versus CuPLA thickness. Additional field trimmers (5-mm thick 92-95% tungsten-doped PLA) are used to reduce out-of-field transmission. The compensator and trimmer models are 3D printed using a fused deposition modeling type printer. Each compensator and trimmer are secured in a custom treatment cone that preferentially attenuates the uniform x-ray beam to achieve a uniform dose to the tumor with a sharp dose reduction to the adjacent healthy tissue using the IMRT plan developed.

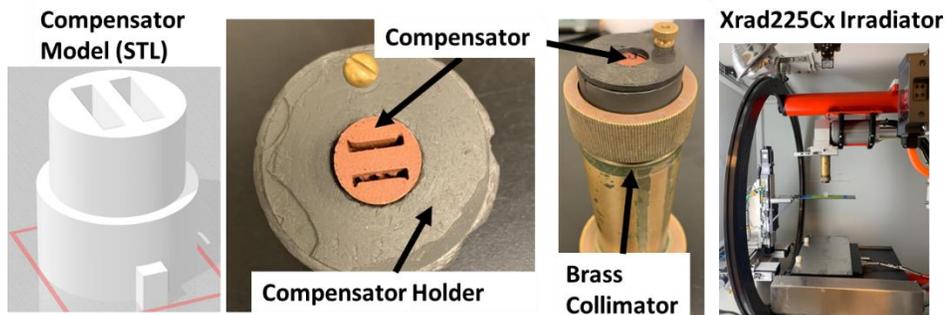

**Figure 1:** SA-IMRT system. A 3D compensator model (left) is generated from optimized fluence patterns. The printed compensators are inserted in a custom brass collimator (center) and attached to a conformal small animal irradiator (right) prior to treatment delivery.

### 2.B. Animal subject contouring and hypoxia imaging

A total of seventeen mice, each with a leg born fibrosarcoma (FSa) tumor were considered in this study. Mouse data were acquired under IACUC protocol 71697. Each animal had a set of rigidly registered CT, MRI, and EPRI oxygen volumes. A tumor contour was generated from the T2-weighted MRI volume. A hypoxic target volume (HTV) was defined by thresholding tumor voxels with oxygen concentrations less than or equal to 10 mmHg measured by EPRI. Each animal had one to six discrete hypoxic regions within the primary tumor. The HTV was removed from the tumor contour to formulate a PTV (*PTV = tumor – HTV*) and avoid contradicting optimization objectives during treatment planning. Additional contours corresponding to non-tumor leg tissue (body), bone, and immobilization devices were manually contoured from the CT image set using 3D Slicer. A skin contour was defined by eroding the body contour by 1 mm. PTV and HTV voxels within the skin volume were removed to account for non-target hair follicles in the animal fur that may be classified as hypoxic. Finally, the native CT resolution of 0.1 mm was scaled to 0.2 mm to reduce computation time for dosimetric calculations performed on an equivalent grid size.

### 2.C. Treatment planning strategy

Preclinical radiation treatment plans were generated using IMRT and CRT techniques. For each animal, plans were generated to deliver a uniform 22.5 Gy to the tumor PTV ($Rx_{PTV}$). A simultaneous integrated boost of 13 Gy was prescribed to the HTV resulting in a total dose of 35.5 Gy ($Rx_{HTV}$). The total number of beam angles, distributed uniformly over 360 degrees, was varied as 2, 3, 5, 7, 9, and 15. For the CRT plans, two sets of field apertures were used at each beam angle. The first set of field apertures were configured to conform to the union of PTV and HTV. The second set of field apertures were configured to conform to the HTV. A 0.9 mm block margin was applied to the PTV and HTV.

Automated treatment planning was performed using the inverse TPS described in Section 2.A with the default MatRad IPOPT optimizer and convergence criteria (tolerance = 1e-8, 500 maximum iterations). For a given plan, the three-dimensional dose distribution, **d**, was determined by finding the fluence pattern, **x**, which minimized a weighted piecewise cost function, f(**x**), as shown in Equation 1. Here, f(**x**) is the sum of *n* individual objective functions $f_n(x)$ and $p_n$ is the scalar weighting factor for the $n^{th}$ objective function. The total dose to a specific voxel indexed by *i* was computed as $d_i = \sum_j D_{ij} x_j$, where **D** is the dose influence matrix and $D_{ij}$ defines the dose contribution of bixel *j* with weight $x_j$ to voxel *i*. Bixel widths of 0.5 mm and 0.25 mm at isocenter were used for the PTV and HTV fields, respectively, for the CRT technique. Due to computational considerations, a fixed bixel resolution of 1.0 mm at isocenter was used for IMRT.

$$\min_{x \in \mathbb{R}^B} f(x) = \sum_n p_n f_n(x) \qquad subject\ to\ x \geq 0 \qquad (1)$$

The objective function, *f(x)*, used for IMRT planning in this work consisted of the four terms presented in equation 2.

$$f(x) = \frac{0.4}{N_{PTV}} \sum_{i \in PTV} (d_i - 22.5)^2 + \frac{0.45}{N_{HTV}} \sum_{i \in HTV} (d_i - 35.5)^2 + \frac{0.10}{N_{Ring}} \sum_{i \in Ring} (d_i - 22.5)^2$$
$$+ \frac{1}{N_{Body}} \sum_{i \in Body} \Theta(d_i - 0.5 \cdot 22.5)(d_i - 0.5 \cdot 22.5)^2 \qquad (2)$$

Here, $\Theta(x)$ denotes the Heaviside function defined as: $\Theta(x) = 0\ if\ x < 0\ and\ \Theta(x) = 1\ if\ x \geq 0$.

The first term ensures the prescription dose is delivered uniformly to the PTV. The second term encourages a uniform dose of 35.5 Gy within the HTV. The third term is used to tightly conform dose around the HTV boost region using a ring planning structure. The ring planning structure was generated by expanding the hypoxic sub-volume 0.9 mm within the PTV and then removing the HTV from the expanded structure. Finally, the fourth term is a square over-dosage function that penalizes dose to non-target structures generically termed herein as *body.* The same objective function and weights were used for each combination of beam angles considered. Forward planning with evenly weighted beams was performed for the CRT technique. Following optimization, each plan was normalized such that 95% of the HTV received the 35.5 Gy prescription dose. As a result, some variation in the reported D95% values of the PTV for which 22.5 Gy was prescribed but not re-normalized is expected.

## 2.D. Evaluation of treatment plan quality

Treatment plan quality was evaluated in terms of dose conformity about the HTV, dose heterogeneity within the PTV, and agreement with the treatment planning goals assessed in terms of several dose volume histogram metrics and statistics.

Dose conformity was specified in terms of the Paddick Conformity Index (CI) defined by Equation 3.[37]

$$CI_{HTV} = \frac{(HTV_{PIV})^2}{HTV \cdot PIV} \quad (3)$$

Here, $HTV_{PIV}$, is the hypoxic target volume covered by the hypoxic prescription isodose volume, HTV is the hypoxic target volume, and PIV is the total volume covered by the hypoxic prescription isodose. $CI_{HTV}$ has an ideal value of 1.0 and conformity is indicated to worsen with a decreasing index value.

The spread between the differential dose volumes was characterized by the D25% difference (minimum dose to 25% of the volume of interest) and D50% difference of the HTV and PTV (HTV-PTV) volumes. A difference of 13 Gy would correspond to an ideal step function between the HTV and PTV.

Dose uniformity within the PTV is measured in terms of the mean, standard deviation, and ratio of the standard deviation to mean dose (σ/μ) within the PTV. This metric was utilized instead of the more common homogeneity index defined as $HI = \frac{D_{max}}{D_{Rx}}$ to reduce the sensitivity of the metric to high dose values immediately adjacent to the hypoxic boost region. Smaller values of (σ/μ) indicate a more uniform dose within the volume-of-interest while higher values indicate greater dose heterogeneity.

The extent to which a plan satisfies the planning goals is quantified using the D95% (dose to 95% of the volume of interest), mean dose, and standard deviation of dose within the PTV and HTV. This is further characterized by presenting dose volume histograms.

The statistical significance of plan quality metrics for IMRT versus CRT was evaluated using a paired, two-sided Wilcoxon signed-rank test with Bonferroni correction to account for the multiple comparisons performed across six different sets of beam angles. A p-value less than or equal to 0.05 was considered significant; a p-value less than or equal to 0.001 was considered highly significant.

## 2.E. Evaluation of target shape complexity

To determine whether there could be a subset of animals presenting with complex distributions of hypoxia for which IMRT may be particularly advantageous, correlations between IMRT and CRT plan quality in terms of $CI_{HTV}$ versus hypoxic target shape complexity were determined. The surface-area-to volume ratio and number of discrete hypoxic regions were used as the metrics of hypoxic target complexity. Surface-area-to-volume was computed as the volume weighted ratio of the HTV surface area to HTV volume. The surface-area-to-volume ratio increases as target spiculatedness increases. A linear least-squares fit was performed. The Pearson correlation coefficient and coefficient of determination ($R^2$) was computed.

### *2.F. Selection of minimum number of required beam angles*

Compensator fabrication times, filament material, and treatment delivery times increase as more beam angles are used. To determine the minimum number of beam angles required to balance plan quality versus treatment delivery times for preclinical dose painting of tumor hypoxia in a mouse flank model, the D50% and D25% difference metrics were evaluated as a function of number of beam angles and tested for significance versus the 3-angle base case using a paired, two-sided Wilcoxon signed-rank test with Bonferroni correction and a 0.05 significance level.

## 3. RESULTS

### *3.A. Dose painting tumor hypoxia*

Figure 2 presents IMRT and CRT dose distributions and differences for two example plans. Qualitatively, improved dose conformity and reduced dose spillage around the HTV is observed for IMRT versus CRT. A greater separation of the high dose volume prescribed to the HTV (35.5 Gy) versus the lower dose prescribed to the PTV (22.5 Gy) is apparent.

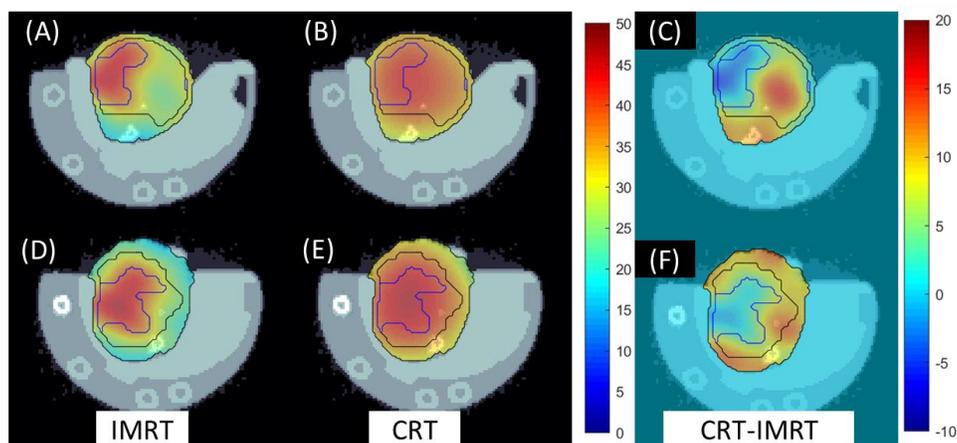

**Figure 2:** Dose distributions shown for IMRT (A, D) show improved conformity about the HTV contour shown in blue relative to CRT (B, E) for two example animals (top versus bottom rows). Dose differences (CRT minus IMRT) are presented in (C, F).

Figure 3 presents box plots of the D25% and D50% differences, respectively, for IMRT and CRT plans as a function of the number of beam angles to quantify the dosimetric spread between the uniformly targeted PTV and the dose painted HTV. D25% and D50% differences are presented for each individual animal, along with summary statistics, in Table 1 for the 3-angle treatment plans. Examination of the individual data points in Figure 3 shows the outliers most corresponded to animals with a high number (≥ 5) of discrete hypoxic regions.

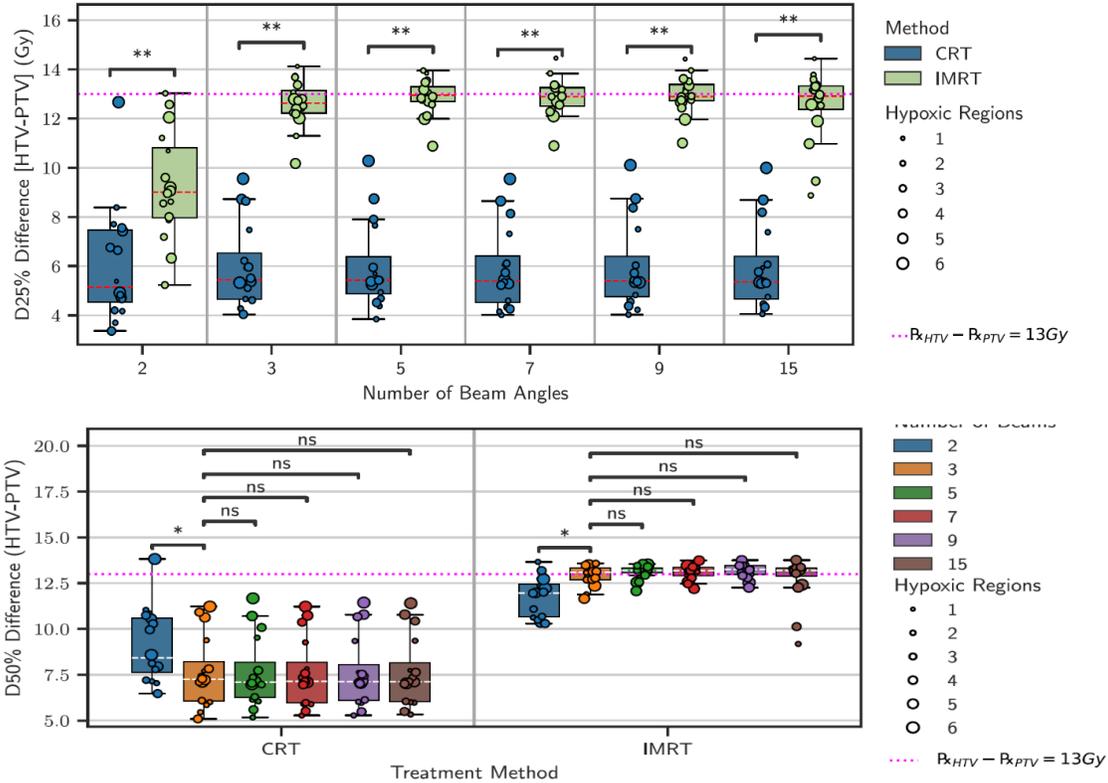

**Figure 3:** The D25% (top row) and D50% (bottom row) differences between the HTV and PTV for IMRT and CRT techniques are shown versus the ideal 13 Gy difference line. IMRT D25% differences were significantly ($p < 0.001$ annotated by **) improved relative to the CRT technique. Use of 3-angles significantly ($0.001 < p \leq 0.05$, annotated by *) improved D50% differences for IMRT and CRT. No significant ($p > 0.05$) differences were observed as the number of beam angles increased beyond 3. Red dashed lines indicate median values, box edges indicate 1st and 3rd quartiles, and whiskers are set at +/- 1.5 multiplied by the interquartile range. Individual data points are shown for each animal with marker sizes defined as a function of the total number of discrete hypoxic regions present.

**Table 1:** Plan quality metrics are summarized for IMRT and CRT for 3-angle treatment plans.

| Case ID | CI (HTV) IMRT | CI (HTV) 3DCRT | D25% Difference IMRT | D25% Difference 3DCRT | D50% Difference IMRT | D50% Difference 3DCRT | PTV Dose Mean ± σ (Gy) IMRT | PTV Dose Mean ± σ (Gy) 3DCRT | PTV Uniformity (σ/mean) IMRT | PTV Uniformity (σ/mean) 3DCRT | PTV D95% Dose (Gy) IMRT | PTV D95% Dose (Gy) 3DCRT |
|---|---|---|---|---|---|---|---|---|---|---|---|---|
| 1 | 0.46 | 0.16 | 13.8 | 6.6 | 13.2 | 8.3 | 25.7 ± 2.2 | 29.1 ± 3.6 | 8.7% | 12.5% | 22.5 | 22.4 |
| 2 | 0.64 | 0.24 | 14.1 | 5.9 | 13.4 | 7.7 | 26.8 ± 2.6 | 31.6 ± 4.5 | 9.6% | 14.3% | 23.2 | 22.8 |
| 3 | 0.50 | 0.29 | 12.4 | 4.3 | 12.9 | 5.5 | 28.8 ± 3.0 | 33.8 ± 4.8 | 10.4% | 14.2% | 24.5 | 24.8 |
| 4 | 0.43 | 0.12 | 13.1 | 7.5 | 13.1 | 9.4 | 26.8 ± 2.7 | 32.1 ± 5.0 | 10.2% | 15.6% | 23.3 | 23.9 |
| 5 | 0.39 | 0.17 | 12.8 | 5.3 | 13.5 | 7.1 | 29.1 ± 3.3 | 33.4 ± 4.7 | 11.4% | 14.0% | 24.5 | 25.0 |
| 6 | 0.29 | 0.14 | 10.2 | 5.4 | 11.7 | 7.2 | 29.0 ± 3.7 | 32.9 ± 4.6 | 12.6% | 14.0% | 23.9 | 24.2 |
| 7 | 0.34 | 0.14 | 12.5 | 5.1 | 13.1 | 7.1 | 29.5 ± 3.4 | 34.1 ± 5.1 | 11.5% | 15.1% | 24.8 | 25.3 |
| 8 | 0.48 | 0.23 | 12.2 | 5.5 | 12.7 | 7.3 | 28.4 ± 3.2 | 32.6 ± 4.9 | 11.3% | 15.0% | 24.0 | 23.2 |
| 9 | 0.58 | 0.23 | 13.7 | 4.3 | 13.3 | 6.0 | 27.6 ± 2.6 | 32.9 ± 5.2 | 9.5% | 15.8% | 24.3 | 23.2 |
| 10 | 0.24 | 0.05 | 12.4 | 8.7 | 13.1 | 10.6 | 28.8 ± 3.5 | 37.0 ± 5.4 | 12.0% | 14.6% | 23.9 | 27.7 |
| 11 | 0.42 | 0.16 | 12.8 | 6.0 | 13.1 | 7.8 | 28.2 ± 3.1 | 32.7 ± 4.3 | 11.1% | 13.3% | 23.6 | 24.8 |
| 12 | 0.26 | 0.04 | 13.4 | 8.6 | 13.5 | 10.9 | 27.2 ± 2.6 | 32.8 ± 4.7 | 9.7% | 14.3% | 23.5 | 24.0 |
| 13 | 0.55 | 0.19 | 12.2 | 4.6 | 12.7 | 6.1 | 27.5 ± 3.2 | 32.9 ± 4.1 | 11.7% | 12.6% | 22.6 | 24.3 |
| 14 | 0.58 | 0.33 | 12.7 | 4.0 | 12.8 | 5.1 | 29.2 ± 3.2 | 34.2 ± 4.5 | 11.1% | 13.1% | 24.4 | 25.1 |
| 15 | 0.45 | 0.26 | 11.3 | 4.7 | 11.9 | 5.8 | 30.0 ± 3.3 | 34.1 ± 4.7 | 11.0% | 13.8% | 25.3 | 24.9 |
| 16 | 0.65 | 0.26 | 13.7 | 6.2 | 13.6 | 7.4 | 28.1 ± 2.9 | 35.4 ± 5.3 | 10.3% | 14.9% | 23.9 | 25.1 |
| 17 | 0.06 | 0.01 | 12.0 | 9.6 | 12.4 | 11.2 | 25.6 ± 2.4 | 28.5 ± 4.1 | 9.4% | 14.3% | 22.6 | 23.1 |
| Median | 0.45 | 0.17 | 12.7 | 5.5 | 13.1 | 7.3 | 28.2 | 32.9 | 11.0% | 14.3% | 23.9 | 24.3 |
| Mean | 0.43 | 0.18 | 12.7 | 6.0 | 12.9 | 7.7 | 28.0 | 32.9 | 10.7% | 14.2% | 23.8 | 24.3 |
| σ | 0.16 | 0.09 | 1.0 | 1.7 | 0.5 | 1.9 | 1.3 | 32.9 | 1.1% | 1.0% | 0.8 | 1.3 |
| p-value | 1.8E-04 | | 1.8E-04 | | 1.8E-04 | | | | 1.8E-04 | | 3.0E-01 | |

## 3.B. IMRT versus CRT dose conformity

Figure 4 presents $CI_{HTV}$ for IMRT versus CRT. The IMRT dose distributions were significantly (p<0.001) more conformal for each number of beam angles considered. For the 3-angle scenario, median $CI_{HTV}$ was 0.45 with IMRT versus 0.17 for CRT.

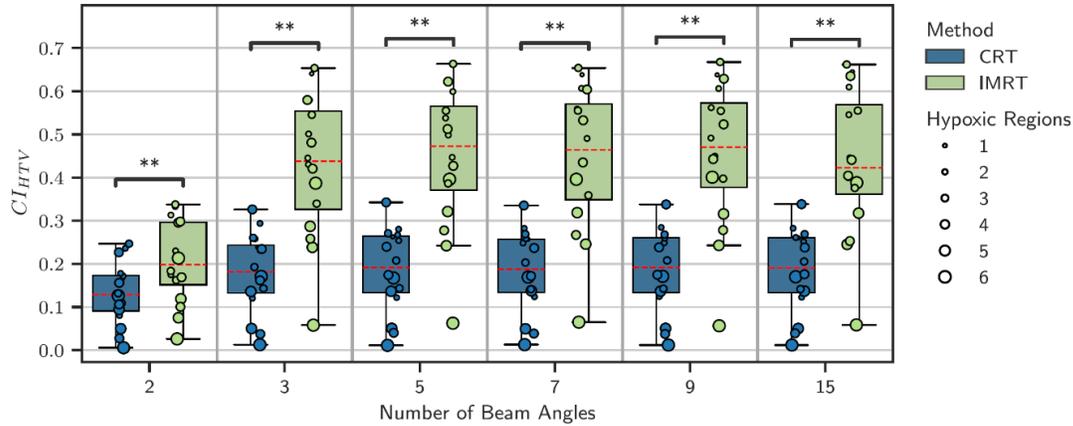

**Figure 4:** The conformity index is shown for IMRT and CRT techniques. Dose conformity about the HTV was significantly improved with use of IMRT versus CRT.

## 3.C. Optimal number of beam angles for dose painting tumor hypoxia

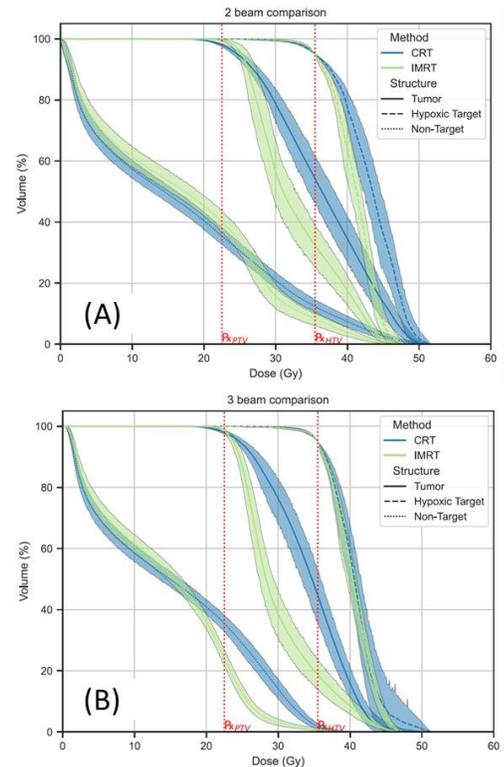

The preceding two sections (3.A. and 3.B.) demonstrated the benefits of IMRT versus CRT for preclinical dose painting with an equivalent number of beam angles. Figure 3 presents D50% differences with statistical significance tested within each technique as a function of the number of beam angles versus the 3-angle scenario. Differential SIB dosing, in terms of D50%, was significantly improved with the use of three beam angles for both IMRT and CRT relative to the two-angle case. This is further supported by DVH curves presented in Figure 5 panel (A) for 2-angle treatments and panel (B) for 3-angle treatments. Qualitatively one notices increased separation between the HTV and PTV curves with use of 3-angles versus 2-angles. No noticeable impact was observed from 3-angles to 5-angles (Supplemental Figure 1). The median D25% and D50% differences increased from 9.1 Gy and 12.0 Gy, respectively, to 12.7 Gy and 13.1 Gy for 2-angle and 3-angle IMRT. Median D25% and D50% differences changed from 5.4 Gy and 8.3 Gy to 5.5 Gy and 7.3 Gy for CRT. No significant (p>0.05) improvement in D50% difference was observed as the number of beam angles increased from 3 angles to 5, 7, 9 or 15 (Figure 3). Thus, for the purposes of this work we assumed 3-angle IMRT was sufficient for dose painting tumor hypoxia in a murine flank model. It's acknowledged, however, that a significant improvement (p=0.034) in IMRT D25% difference was observed as the number of beam angles increased from 3 to 9. While statistically significant, the absolute mean difference of 0.3 Gy (12.7 Gy increased to 13.0 Gy) should be considered versus the additional material cost and delivery time required to use nine beam angles.

**Figure 5:** Dose volume histograms are shown for IMRT and CRT performed with 2-angles (A) and 3-angles (B).

Additionally, no significant improvement in D25% was observed as the number of beam angles increased from 3 to 15.

### 3.D. Radiobiological considerations: target coverage and dose uniformity

As stated previously, each of the treatment plans was normalized such that the D95% to the HTV equaled the prescription dose of 35.5 Gy. The D95% to the PTV averaged 23.8 [range of 22.5 to 25.3] Gy for IMRT and 24.3 [range of 22.4 to 27.7] Gy for CRT, as shown in Table 1, versus the 22.5 Gy goal. No significant differences in D95% between IMRT and CRT were observed for treatment plans using 3 or more beam angles. In contrast, significant improvements in PTV dose uniformity were observed for IMRT versus CRT plans for a fixed number of beam angles. Figure 6Figure 6 presents the PTV dose uniformity metric (σ/μ) for each technique and number of beam angles. No significant improvements in PTV dose uniformity were observed for CRT with the use of additional beam angles relative to the three-angle plans. For IMRT, significant improvements in dose uniformity were observed as the number of beams was increased.  This is supported by average σ/μ decreasing from 10.7% with 3 beam angles, to 10.2% (5-angles), 10.1% (7-angles), 10.1% (9-angles), and 9.9% (15 angles).

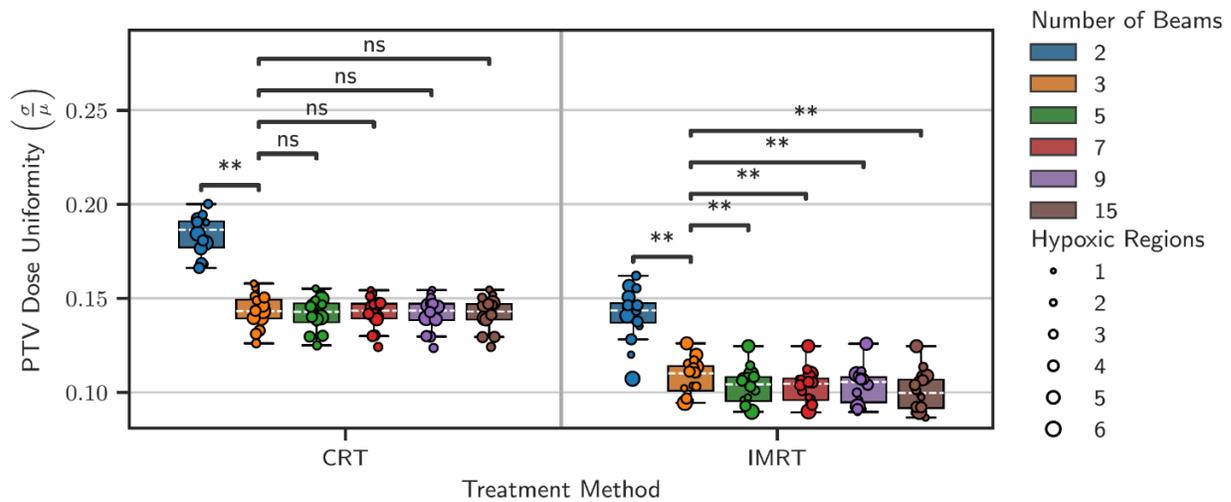

**Figure 6:** PTV dose uniformity is presented for CRT and IMRT techniques. For CRT, no significant differences in dose uniformity were observed as the number of beam angles increased beyond 3. For IMRT, significant improvements in dose uniformity were observed with an increased number of beams.

### 3.E. Plan quality versus target complexity

Figure 7 presents the linear regressions of three-angle IMRT and CRT plan quality assessed in terms of conformity index versus target shape complexity metrics, surface-area-to-volume ratio (A) and number of hypoxic regions (B). The slope, intercept, $R^2$, and correlation coefficient values (ρ) are included for each regression. Pearson correlation coefficient values for $CI_{HTV}$ versus surface-area-to-volume ratio were -0.89 for IMRT and -0.93 for CRT, respectively, when excluding a single case deemed an outlier. Including the outlier case, which contained six discrete hypoxic target regions and the maximum observed surface-area-to-volume ratio, decreased the correlation coefficients to -0.83 and -0.75, respectively.

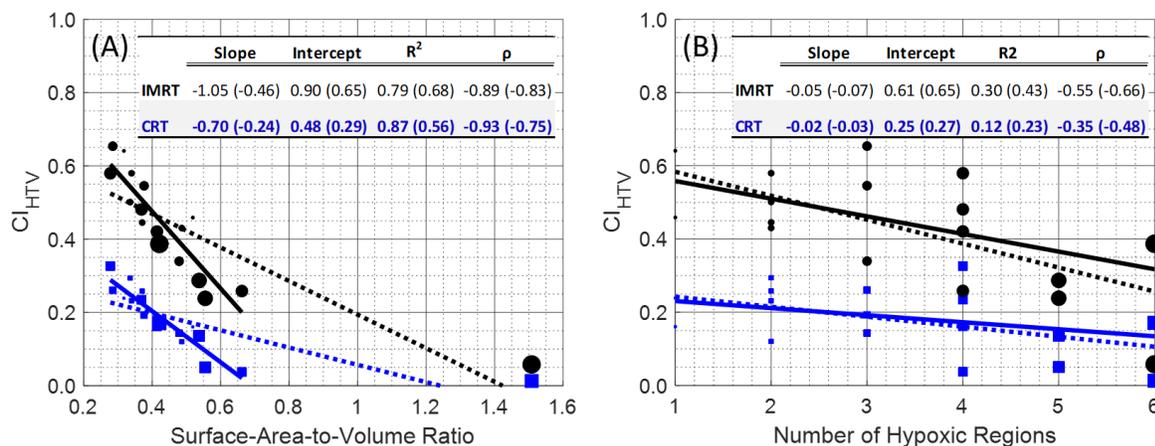

**Figure 7:** Plan quality in terms of conformity index is plotted versus target shape complexity in terms of surface-area-to-volume radio (A) and number of hypoxic regions (B) for IMRT and CRT plans. Marker sizes correspond to the number of discrete hypoxic regions present for each case. Linear regression parameters are displayed. Values in parentheses (and dotted lines) correspond to an analysis including all 17 animals, while values without parentheses (solid lines) exclude case 17 which appeared to be an outlier.

### 4. DISCUSSION

The primary aim of this work was to investigate the dosimetric consequences of applying SA-IMRT for radiation dose painting in a preclinical setting using a novel compensator-based delivery system. While several methods and systems to perform intensity modulated radiotherapy for small animal subjects have recently been proposed, no systematic evaluations have been performed comparing the dosimetric properties of SA-IMRT versus CRT using real animal data. In this study, we quantified differences in IMRT versus CRT radiation plans designed to dose paint tumor hypoxia identified from EPRI in seventeen animal subjects with fibrosarcoma tumors.

Our results presented in Figure 3 demonstrate that IMRT provides statistically *significant* ($p<0.001$) improvements in selectively boosting (via an SIB approach) dose to the hypoxic tumor volume regions relative to CRT for each number of beam angles considered. The median value of the D25% dose differences between the HTV boost volume and PTV volume was 9.1 Gy with 2-beam angles, and approached or met the ideal 13 Gy dose difference as the number of beam angles was increased to 3 (12.7 Gy), 5 (13.0 Gy), 7 (12.9 Gy), 9 (12.9 Gy), and 15 (13.0 Gy), using our in-house preclinical SA-IMRT technique. Median D25% differences were smaller and measured 5.4 Gy, 5.5 Gy, and 5.4 Gy, for 2-, 3-, and 5 to 15- beam angles respectively, using the CRT technique. Trends were similar and statistically significant for D50% differences also. As a result, we expect that IMRT will improve current preclinical radiation delivery capabilities to precisely target and differentiate dose to hypoxic tumor regions versus non-hypoxic regions. We consider this to be an important technological contribution that could enable preclinical studies that leverage advanced molecular imaging techniques to preferentially target tumor sub volumes and probe tumor heterogeneities to produce findings more readily translatable to the clinical setting.

The increased separation in dose administered to the HTV versus the PTV with IMRT may be attributed to significant improvements in dose conformity ($CI_{HTV}$) about the HTV as presented in Figure 4. These results demonstrate that SA-IMRT could be deployed to better differentiate dose levels to multiple target volumes as is required for dose painting tumor hypoxia. However, a large spread in $CI_{HTV}$, visualized in terms of the interquartile range, and Table 1 values is apparent. This suggests some distributions of hypoxia may be well suited for IMRT treatments while others may be too complex to fully benefit from the added complexity of the delivery technique.

To assess the relationship between hypoxic target shape complexity and treatment plan quality, a series of linear regressions were performed (Figure 7). As expected, the $CI_{HTV}$ was inversely correlated and degraded with increased target complexity assessed in terms of the surface-area-to-volume ratio and number of discrete hypoxic target regions. The derived plan quality and target shape complexity relationships could potentially be used to identify animals with spatial distributions of hypoxia that are well suited for IMRT. For example, the IMRT $CI_{HTV}$ was reduced to half of the maximum observed value (0.65) for surface-area-to-volume ratios greater than 0.54 using the reported linear regression equations.

In addition to superior dose conformity and differential dosing capabilities, IMRT plans were associated with improved dose uniformity versus CRT plans (Figure 6). Improved dose homogeneity with IMRT versus CRT is a potentially important finding relevant to radiobiological studies focused on evaluating steep-dose response relationships. While this work concluded 3-angles offered a reasonable compromise between plan quality (D25% difference) for dose painting tumor hypoxia versus increased delivery time and complexity, the results presented here suggest the optimal number of beam angles required may differ for the applications that prioritize dose uniformity. Future planned studies will quantify the potential improvements in single target (i.e., non dose painting application) uniformity for IMRT versus CRT.

Workflow efficiency and animal throughput are important considerations in preclinical experiments. The results of this work found that 3-angle IMRT is sufficient for dose painting tumor hypoxia in a murine flank model when plan quality is assessed in terms of the D50% difference. The use of less beam angles would reduce the overall compensator fabrication time, treatment delivery time, and total amount of printing filament required (i.e., cost). Although three beam angles were found to be optimal for the task of dose painting tumor hypoxia, additional evaluation is required for other radiobiological applications. A significant improvement in $CI_{HTV}$ was observed with use of 5, 7, and 9 beam angles relative to the 3-angle scenario (supplemental figure 2). Applications that require greater than 3 beam angles may benefit from fabricating beam compensators in parallel using multiple 3D printers. Finally, a limitation of this work is that beam angles were uniformly distributed over a 360-degree arc. Future work will investigate sparse optimization techniques to optimally select beam orientations while simultaneously reducing the total number of required beam angles.[38] Optimal selection of beam orientations coupled with techniques for fluence field regularization could further reduce the compensator fabrication times and material to improve efficiency.[39]

It's important to highlight that the CRT approach used here offers an additional level of conformality that is not possible with current commercially available fixed-shaped apertures. That is, current commercial systems are limited to using apertures of circular or rectangular shapes while the CRT apertures considered in this paper were designed to conform to target volumes within the beams-eye-view similar to what is used in clinical settings. As a result, we expect the improvements in plan quality observed for SA-IMRT to be even greater when compared to what could be possibly delivered using standard fixed field apertures. An additional benefit of using the proposed SA-IMRT technique is that only a single compensator is required at each beam angle to perform dose painting. With the CRT approach, each beam angle requires two sets of apertures per beam angle. The first set of apertures are required to target the primary tumor while the second set is designed for boosting the smaller HTV. This is a limitation of CRT as it would potentially require increased fabrication times and is inefficient as each aperture needs to be manually attached to the brass collimator prior to radiation delivery. An additional CRT consideration is that this work used uniformly weighted beam angles. This was done to mimic a forward planning scenario as is most typically done. The results presented here may be altered if CRT beam angles were weighted differently/optimally.

A limitation of this study is that we focused solely on a dose painting application. As a result, our planning objectives and weights emphasized high dose gradients to achieve differential dosing objectives and high target dose conformity. Single target applications focused on organ sparing or dose uniformity may be of interest to some investigators and will be the focus of a future planned study. Nonetheless, this highlights the increased flexibility SA-IMRT provides to pursue different planning objectives that may not be possible using what is currently available in preclinical setting and even with the CRT technique described herein. A second limitation of this study is that we did not perform delivery quality assurance for the individual treatment plans. However, our prior work demonstrated that SA-IMRT plans targeting tumor hypoxia could be accurately delivered with per-field gamma analysis pass rates of 98.8% with 3%/1.0 mm criteria.[33]

## 5. CONCLUSION:

Recent advancements have been made in the development of preclinical irradiation methodologies, including IMRT, that more closely mimic clinical techniques aimed at improving the translation of preclinical results to clinical outcomes. This study investigated the dosimetric implications of employing SA-IMRT in comparison to CRT to administer a simultaneous integrated boost dose of radiation to hypoxic tumor voxels. The results demonstrated improved dose conformity, and increased dose uniformity. IMRT was associated with greater dose separation between hypoxic and normoxic target volumes versus CRT, demonstrating the potential of SA-IMRT to facilitate dose-painting studies based on molecular imaging signals. Future work will include outcome-based studies of oxygen guided intensity modulated radiation therapy in a murine model.

**DATA AVAILABILITY:** All data generated or analyzed during this study are included in this published manuscript (and its Supplementary Information files) or are available from the corresponding author on reasonable request.

**ACKNOWLEDGMENTS:**
This work was supported by funding from the American Cancer Society IRG-19-136-59-IRG-02 grant via the University of Chicago Comprehensive Cancer Center and a Badger Challenge grant from the University of Wisconsin - Madison. Mouse data were acquired under IACUC protocol 71697 and with funding support from National Institutes of Health grants P41 EB002034, R01 CA098575, R50 CA211408. The authors would like to thank Mícheál John Ócolla for assistance with contouring the animal support structures.

**AUTHOR CONTRIBUTIONS:** JMS, EP, RT, and GR contributed to all aspects of the work including software implementation and validation, treatment planning, and data analysis. DOV participated in treatment planning and contouring. BA and HJH contributed to study design. HJH and BE performed the animal subject imaging. JMS drafted the original manuscript. All authors edited, reviewed, and approved the final manuscript.

**ADDITIONAL INFORMATION**

**Competing Interests** HJH holds US patent 8,664,955 (US-8664955-B1) and HJH and BE hold US patent 9,392,957 (US-9392957-B1) for aspects of the pO2 imaging technology. HJH and BE are members of a start-up company O2M Technologies, LLC. The other authors state that the research was conducted without any commercial or financial relationships that could be construed as a potential conflict of interest.

**SUPPLEMENTARY MATERIAL:**

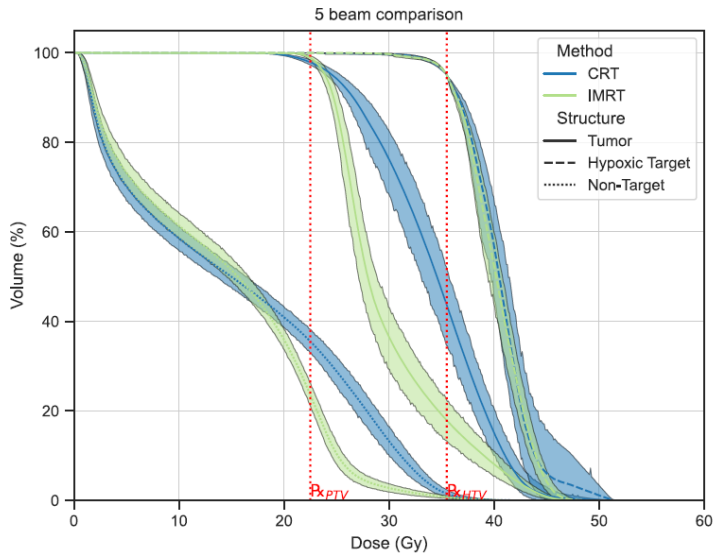

**Supplemental Figure 1:** Dose volume histograms are shown for IMRT and CRT performed with 5-angles.

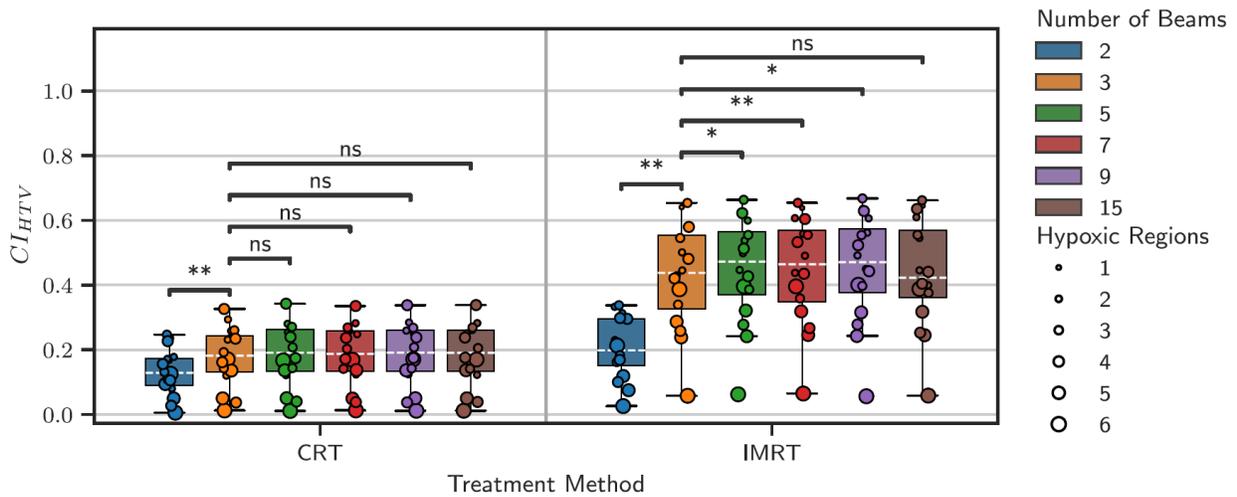

**Supplemental Figure 2:** The conformity index is shown for IMRT and CRT techniques. Statistical significance of number of beam angles is evaluated versus the three-angle case.